\documentclass[reprint, prl, superscriptaddress, floatfix, noeprint]{revtex4-2}

\usepackage{hyperref}
\hypersetup{
    colorlinks=true,
    linkcolor=blue,
    citecolor=red,      
    urlcolor=blue,
    pdftitle={Overleaf Example},
    pdfpagemode=FullScreen,
    }
\usepackage{graphicx}
\usepackage{amsmath}
\usepackage{wasysym}
\usepackage{amsfonts}
\usepackage{color}
\usepackage{verbatim}
\usepackage{upgreek}
\usepackage{bm}
\usepackage{textcomp}
\usepackage[symbol*]{footmisc}
\usepackage{float}
\usepackage{soul}
\usepackage{xcolor}

\DefineFNsymbolsTM{otherfnsymbols}{
  \textdagger    \dagger
  \textdaggerdbl \ddagger
  \textsection   \mathsection
  \textparagraph \mathparagraph
 }
\setfnsymbol{otherfnsymbols}

\newcommand{\beq}{\begin{equation}}
\newcommand{\eeq}{\end{equation}}
\newcommand{\bea}{\begin{eqnarray}}
\newcommand{\eea}{\end{eqnarray}}
\newcommand{\bal}{\begin{align}}
\newcommand{\eal}{\end{align}}

\newcommand{\fig}[1]{Fig.\,\ref{#1}}

\newcommand{\WS}{WS$_2$}

\begin{document}
\title{Tailoring the dielectric screening in \WS\ - graphene heterostructures}

\author{David Tebbe} 
\author{Marc Sch\"utte} 
\affiliation{2nd Institute of Physics and JARA-FIT, RWTH Aachen University, 52074 Aachen, Germany}
\author{Kenji Watanabe}
\affiliation{Research Center for Functional Materials, National Institute for Materials Science, 1-1 Namiki, Tsukuba 305-0044, Japan}
\author{Takashi Taniguchi}
\affiliation{International Center for Materials Nanoarchitectonics, National Institute for Materials Science, 1-1 Namiki, Tsukuba 305-0044, Japan}
\author{Christoph Stampfer}
\affiliation{2nd Institute of Physics and JARA-FIT, RWTH Aachen University, 52074 Aachen, Germany}
\affiliation{Peter Gr\"unberg Institute (PGI-9) Forschungszentrum J\"ulich, 52425 J\"ulich, Germany}
\author{Bernd Beschoten} 
\affiliation{2nd Institute of Physics and JARA-FIT, RWTH Aachen University, 52074 Aachen, Germany}
\affiliation{JARA-FIT Institute for Quantum Information, Forschungszentrum J\"ulich GmbH and RWTH Aachen University, 52074 Aachen, Germany}
\author{Lutz Waldecker} 
\email{waldecker@physik.rwth-aachen.de}
\affiliation{2nd Institute of Physics and JARA-FIT, RWTH Aachen University, 52074 Aachen, Germany}

\begin{abstract}

The environment contributes to the screening of Coulomb interactions in two-dimensional semiconductors. 
This can potentially be exploited to tailor material properties as well as for sensing applications.
Here, we investigate the tuning of the band gap and the exciton binding energy in the two-dimensional semiconductor \WS\ via the external dielectric screening. 
Embedding \WS\ in van der Waals heterostructures with graphene and hBN spacers of thicknesses between one and 16 atomic layers, we experimentally determine both energies as a function of the \WS-to-graphene interlayer distance and the charge carrier density in graphene.
We find that the modification to the band gap as well as the exciton binding energy are well described by a one-over-distance dependence, with a significant effect remaining at several nm distance, at which the two layers are electrically well isolated.
This observation is explained by a screening arising from an image charge induced by the graphene layer.   
Furthermore, we find that the effectiveness of graphene to screen Coulomb interactions in nearby \WS\ depends on its doping level and can therefore be controlled via the electric field effect.
We determine that, at room temperature, it is modified by approximately 20\% for charge carrier densities of $2\times10^{12}$ cm$^{-2}$.

\end{abstract}


\maketitle
\date{\today}

\section{Introduction}

Embedding two-dimensional (2D) semiconductors, such as transition metal dichalcogenides (TMDs), into van der Waals heterostructures offers a wide variety of opportunities for tuning material properties and creating functionalities.
These range from the improvement of the homogeneity of light emission \cite{Ajayi2017, Cadiz2017, Raja2019}, the creation of vertical and lateral heterojunctions \cite{Fang2014, Ceballos2014, Raja2017, Ersfeld2020} to the emergence of hybridized electronic states \cite{Wilson2017} and twist-angle dependent phenomena \cite{Wu2019, Seyler2019}. 
Even in the absence of electronic hybridization, the intrinsic opto-electronic properties of the 2D semiconductor strongly depend on the neighboring materials.
In particular, dielectric screening by the environment modifies Coulomb interactions between charge carriers in 2D semiconductors \cite{Chernikov2014}. 
This dielectric screening manifests as a reduction of exciton binding energies compared to monolayers in vacuum \cite{Stier2016b,Raja2017} and is accompanied by a reduction of the  quasiparticle band gap due to a weaker electron self-interaction \cite{Latini2015, Rosner2016, Raja2017, Park2018}.
Controlling the dielectric environment therefore constitutes a way of tailoring material properties, which can be exploited e.g. to induce lateral heterojunctions in 2D semiconductors  \cite{Rosner2016,Utama2019}.
Importantly, it can be applied in addition to more traditional means, such as alloying, straining or doping \cite{Li2014_alloy, Conley2013, Chernikov2015doping}.

So far, modifications to the external screening have mainly been investigated using TMDs in direct contact with various substrate materials \cite{Raja2017, Utama2019, Waldecker2019, Park2021}.
For real applications, however, it might be necessary to both electrically isolate the two layers as well as to obtain a more gradual control over the dielectric screening.

Electrical isolation could, for example, be achieved using thin hexagonal boron nitride (hBN) as spacer layers \cite{Britnell2012, Xu2018}. 
The effect of thin spacer layers on the screening has been investigated using $GW$ calculations, which predict that the quasiparticle band gap of the TMD decreases with a one-over-distance dependence \cite{Noori2019, Riis-Jensen2020}.
These calculations, however, have experimentally not been tested and also do not describe excitonic effects, which dominate the optical spectra of the TMDs. 

Additional tunability of the screening has been suggested to be possible by changing the charge carrier density of graphene \cite{Qiu2019, Riis-Jensen2020}.
This effect promises an in-situ way of modifying the screening and has yet to be demonstrated at room temperature \cite{Xu2021}. 

Here, we provide a comprehensive picture of the dielectric screening in 2D heterostructures at room temperature using the 2D semiconductor \WS, graphene and thin spacer layers of hBN. 
We quantify the tunability of the exciton binding energy as well as the quasiparticle band gap as a function of the distance between \WS\ and graphene as well as of the charge carrier density of graphene. 
Our results demonstrate a wide range of tunability and will allow the design of heterostructures of 2D semiconductors with (locally) well-defined material properties.

\section{Results}

\subsection{Sample design and preparation}

\begin{figure*}[!bth]
    \begin{center}
        \includegraphics[width=0.95\textwidth]{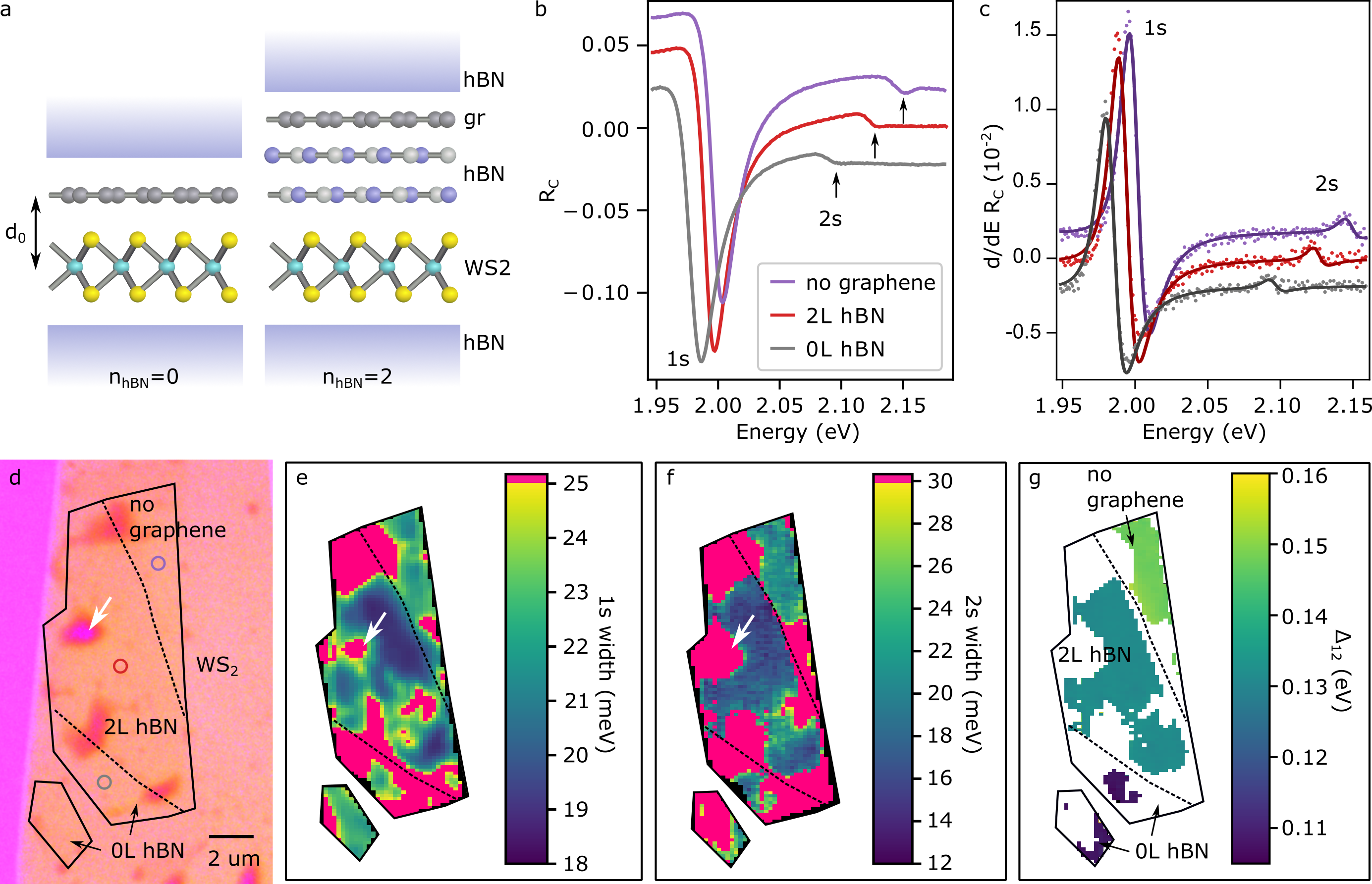}
        \caption{Sample structure and optical measurements. a) Sketch of the 2D heterostructure, consisting of hBN encapsulated \WS\ and graphene monolayers, separated by $n_{\mathrm{hBN}}$ layers of hBN. b) Three typical spectra taken at sample areas with 0 and 2 spacer layers as well as an area with no graphene (positions shown in d). The curves are offset vertically for clarity.  c) Derivative of the spectra (circles) shown in panel b and their respective fits (lines), offset for clarity. d)  Optical microscope image of a typical sample. The black outline denotes the area of \WS\ while the black dotted lines separate areas of different hBN spacer layer thicknesses. e) \& f) Falsecolor maps of the 1s and 2s exciton linewidth of the sample shown in panel d.  g) Map of the 1s-2s exciton level spacing $\Delta_\mathrm{12}$ of the same sample. Areas of increased 1s or 2s linewidths are excluded.}
        \label{fig:imaging}
    \end{center}
\end{figure*}

The samples studied in this work consist of monolayers of \WS\ and graphene, separated by spacers of $n_{\mathrm{hBN}}$ layers of hBN.
These heterostructures are fully encapsulated by thicker top and bottom layers of hBN.
A sketch of the sample geometry with and without spacer layers is shown in \fig{fig:imaging}a.
Some of the stacks contain a graphite gate and an electrical contact to the graphene layer, which allows tuning of the charge carrier density in graphene by a voltage $V_\mathrm{G}$ (see \fig{fig:doping}a). 

All materials were first exfoliated from bulk crystals onto Si/SiO$_2$ wafers. 
Their thicknesses are then characterized by their optical contrasts towards the substrate \cite{Britnell2012}, which were initially calibrated by atomic force microscopy measurements.
The assembly of the stacks was performed using a polymer-based dry transfer technique \cite{Banszerus2015, Pizzocchero2016} in an inert atmosphere of a glovebox (see Methods for details). 

All heterostacks were assembled to contain areas of different thicknesses of hBN spacer layers, and a total of 10 stacks with 26 separate areas were produced in this way.
Since the encapsulating hBN layers are typically more than a 100 layers thick on each side, the effect of any material beyond the outer hBN on the dielectric screening is expected to be negligible \cite{Winther2017}.
Areas in which \WS\ does not directly face a graphene layer therefore serve as a reference, and are treated as heterostacks with infinite layer spacing.

A heterostack containing areas of 0 and 2 hBN  spacer layers as well as an area with no graphene is shown in \fig{fig:imaging}d. 
The different areas are not discernible by their broadband optical contrast in the final stack and are thus marked in the image by dashed lines.

\subsection{Distance dependence of the screening}

We first determine excitonic binding energies and the quasiparticle band gaps of \WS\ in samples of different interlayer spacing. 
Both are obtained from whitelight reflectance contrast spectra at room-temperature via the energy difference of excitonic ground (1s) and excited (2s) states.
This energy is proportional to the exciton binding energy and is used since the energy of the single-particle transitions (the quasiparticle band gap) is not experimentally accessible using optical reflectance spectroscopy \cite{Chernikov2014, Raja2017}.

Typical reflectance contrast spectra are shown in \fig{fig:imaging}b at three positions of different spacer layer thickness of the heterostack shown in \fig{fig:imaging}d (exemplary spectra of every sample are shown in Supplementary Figure 1).
A small red-shift of the excitonic 1s ground state feature (around 2 eV) for thinner spacer layers is visible, accompanied by a much larger red shift of the excited exciton 2s state (marked by arrows). 
Both indicate a simultaneous reduction of exciton binding energy and the quasiparticle band gap with reduced \WS-graphene distance \cite{Raja2017}. 

\begin{figure*}[!tbh]
    \begin{center}
        \includegraphics[width=1.0\textwidth]{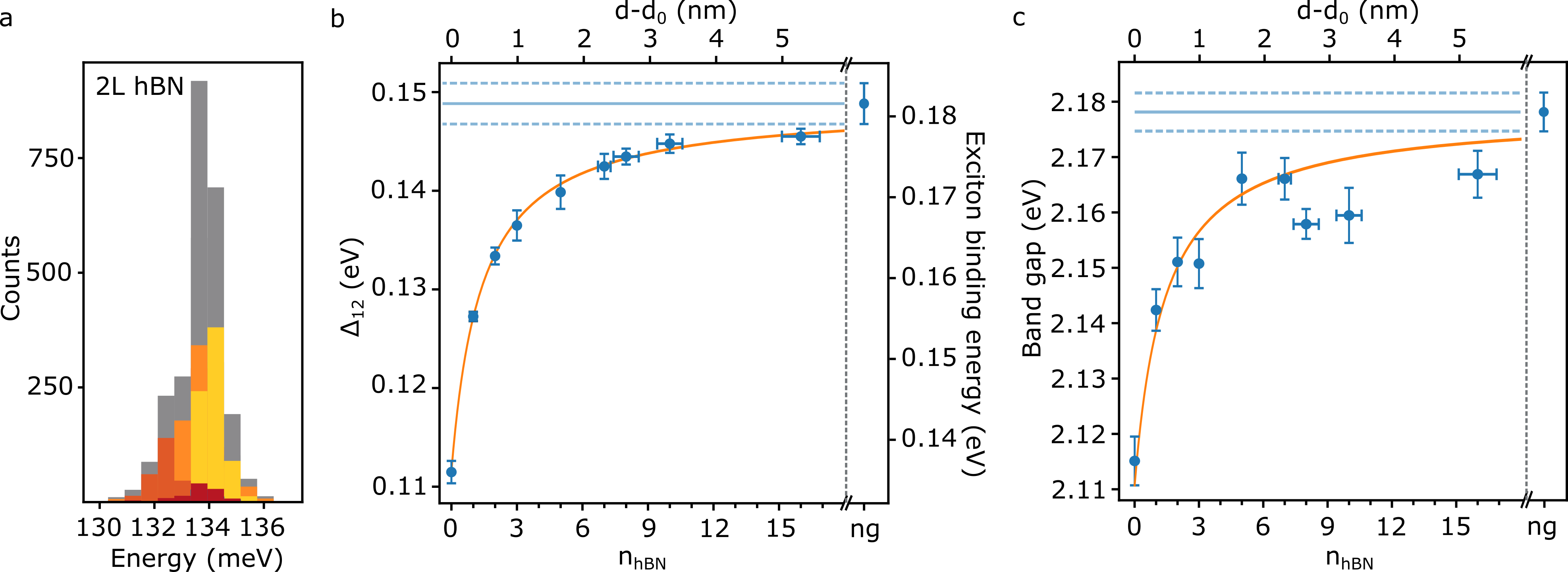}
        \caption{Dielectric screening as a function of \WS-graphene distance. a) Exemplary histogram of the exciton 1s-2s level spacing for two layer thick hBN spacers and four individual samples (total: gray bars, individual samples in color).  
        b) Distance dependence (upper x-axis) of the energy separation of the excitonic 1s and 2s levels (blue points) realized by $n_{\mathrm{hBN}}$ hBN spacer layers (lower x-axis). 
        The approximate exciton binding energy is given in the right axis (see text for details). 
        The data point of no graphene ('ng') is equivalent to an infinite separation of \WS\ and graphene. 
        The orange line is a fit to the data with a one-over-distance dependence. 
        c) Quasiparticle band gap as a function of hBN spacer layers and interlayer distance. The orange line depicts the fit-results using a one-over-distance dependence.
        Datapoints and errorbars in b) and c) are the means and standard deviations of the distribution of values of all samples.         }
        \label{fig:dist}
    \end{center}
\end{figure*}

As the absolute absorption positions can vary between samples and between sample positions e.g. due to local strain \cite{Conley2013}, we recorded reflectance spectra over the entire area of all heterostacks.
The energy of the excitonic absorption features are then determined at each position by fitting the derivative of the measured spectra using a thin-film interference model \cite{Raja2019} (Methods for details).
Examples of the fits and the corresponding data are shown in  \fig{fig:imaging}c. 

Figures \ref{fig:imaging} e \& f depict the color-coded 1s and 2s linewidth of the heterostack shown in Figure \ref{fig:imaging} d.
The linewidth can be taken as an indicator for the local homogeneity of the \WS\ layer, as local strain, doping and dielectric disorder lead to line broadening \cite{Raja2019}. 
We indeed find that areas of increased linewidth are usually accompanied by visible bubbles or folds in the optical micrographs, see e.g. white arrows in \fig{fig:imaging}d-f.  
Note that in large parts of the sample, the linewidth is found to be $\approx 20$ meV, which is slightly below the calculated homogeneous linewidth for \WS\ at room temperature \cite{Selig2016}.
We exclude areas in which the 1s (2s) linewidths exceed 25 (30) meV in the further analysis (indicated in pink in Figures \ref{fig:imaging}\ e \& f). 
\fig{fig:imaging}g shows the 1s-2s separation, $\Delta_{12}$, in the remaining parts of the sample. 
In the image, the three areas of different hBN thicknesses can clearly be distinguished.

We now analyze the distance dependence of the screening in more general terms, for which the above measurement and analysis is repeated on nine additional heterostacks.
For each spacer layer thickness, we obtain a distribution of values of the 1s-2s level spacing.
An example of the distribution of these values, collected on 4 different samples containing a two-layer hBN spacer, is shown in Fig. \fig{fig:dist} a. 
Typically, these distributions have widths of few meV and sample-to-sample variations are of the same order as variations within individual samples. 

The mean value of the distributions of each interlayer distance is shown in \fig{fig:dist} b, with the error bars indicating their standard deviations. 
To account for the smaller standard deviation of histograms to which only a single sample contributes, those errors have been multiplied by a factor of two, conservatively estimated from the sample-to-sample variations of the two-layer area.
For larger hBN thicknesses, the exact determination of hBN layer number becomes more difficult, leading to a small uncertainty in their thickness as well. 
The experimentally determined value for \WS\ with no graphene is shown as a horizontal line (dashed lines indicate the error bars). 

From the 1s-2s level spacing, we calculate the exciton binding energy as $E_B \approx b \cdot \Delta_{12}$.
The proportionality factor depends on the exact sample geometry and is estimated from $GW$ calculations of open-faced samples to be $b=1.2$ \cite{Waldecker2019} (see Methods for details).
Note that the factor $b$ is expected to decrease with increasing screening, such that assuming a constant factor slightly underestimates the change of the binding energy \cite{Waldecker2019}. 
The approximate exciton binding energy is nevertheless given on the right axis of \fig{fig:dist} b. 

The data of $\Delta_{12}$ and the exciton binding energies show a monotonic increase with spacer layer thickness. 
Notably, they remain well below the values for purely hBN encapsulated \WS, even for the thickest spacers investigated here, equalling 16 layers or 5.3 nm \cite{Pease1950}.

In the simplest model, the reduction of Coulomb interactions in the 2D semiconductor arises from the electrostatic interaction of charges in the semiconductor with their image charges, which are situated at twice the distance to graphene. 
In this model, the screening-induced band gap modification is expected to follow a one-over-distance dependence \cite{Neaton2006, Garcia-Lastra2009}.
It has been argued that the model will hold to a good degree for heterostructures of 2D semiconductors and graphene \cite{Riis-Jensen2020}.

We thus fit the data with the relation 
\begin{equation}
\label{eq:dist}
E_B = E_{B,\mathrm{ng}} - \frac{a}{2(d+d_0)},    
\end{equation}
where $E_{B,\mathrm{ng}}$ is the exciton binding energy of \WS\ encapsulated in bulk hBN with no graphene, $d$ is the hBN spacer thickness and $d_0$ is the distance between the center of the TMD and the first graphene or hBN layer, see Fig.~1a, respectively (the two materials have been found to have very similar interlayer distances in heterostructures \cite{Haigh2012}). 
As a material specific parameter, $a$ is proportional to the magnitude of the screening and will be discussed in more detail in the next section.
$E_{B,\mathrm{ng}}, d_0$ and $a$ are left as free parameters in the fit while the hBN layer thickness is given by  $d=n_{\text{hBN}}\cdot3.3$ \AA\ \cite{Pease1950}. 

The fitted curve is shown as a solid orange line in \fig{fig:dist}b with the optimized parameters of $d_0 = (4.7 \pm 0.3)$ \AA\ and $a = (126 \pm 7$) meV\AA. 
The obtained interlayer distance $d_0$ is within the error of the distance of TMDs to hBN obtained from cross-sectional TEM measurements of ($5\pm0.5$) \AA\ \cite{Rooney2017}.

The distance dependence indeed well describes the exciton binding energies in the investigated range of spacer layer thicknesses and the quality of the fit gives high confidence in the validity of the  model. 
It thus demonstrates that it can be applied to excitons, despite them being neutral particles for which multipole screening could lead to deviations.
Such deviations, however, likely only become important at interlayer distances smaller than the thickness of the TMD itself \cite{Waldecker2019}.

From the measurement of the 1s exciton position $E_{1\mathrm{s}}$ and the approximate exciton binding energy, the respective quasiparticle band gap can immediately be calculated as $E_g = E_{1\mathrm{s}}+E_B$, see \fig{fig:dist}c. 
Due to small strain or doping variations, however, the absolute position of the 1s exciton fluctuates much more than the 1s-2s distance, leading to a relatively large error of those data.
Nevertheless, a one-over-distance dependence (with $d_0$ fixed to the previously determined value and $E_{g,\mathrm{ng}}=(2.178 \pm 0.003)$ eV) can also be used to fit the calculated quasiparticle band gap and results in a value of $a=(191 \pm 13)$ meV\AA. 
The larger value of $a$ for the band gap reflects the fact that its absolute change is slightly larger than the one of the exciton binding energy, as also seen in the red shift of the 1s absorption peak. 
This distance-dependence is in qualitative agreement with theoretical calculations but of smaller magnitude \cite{Riis-Jensen2020}, which likely arises due to the absence of a top hBN layer in the calculations. 

\subsection{Tuning the dielectric function of graphene}

\begin{figure*}[!bth]
    \begin{center}
        \includegraphics[width=1.0\textwidth]{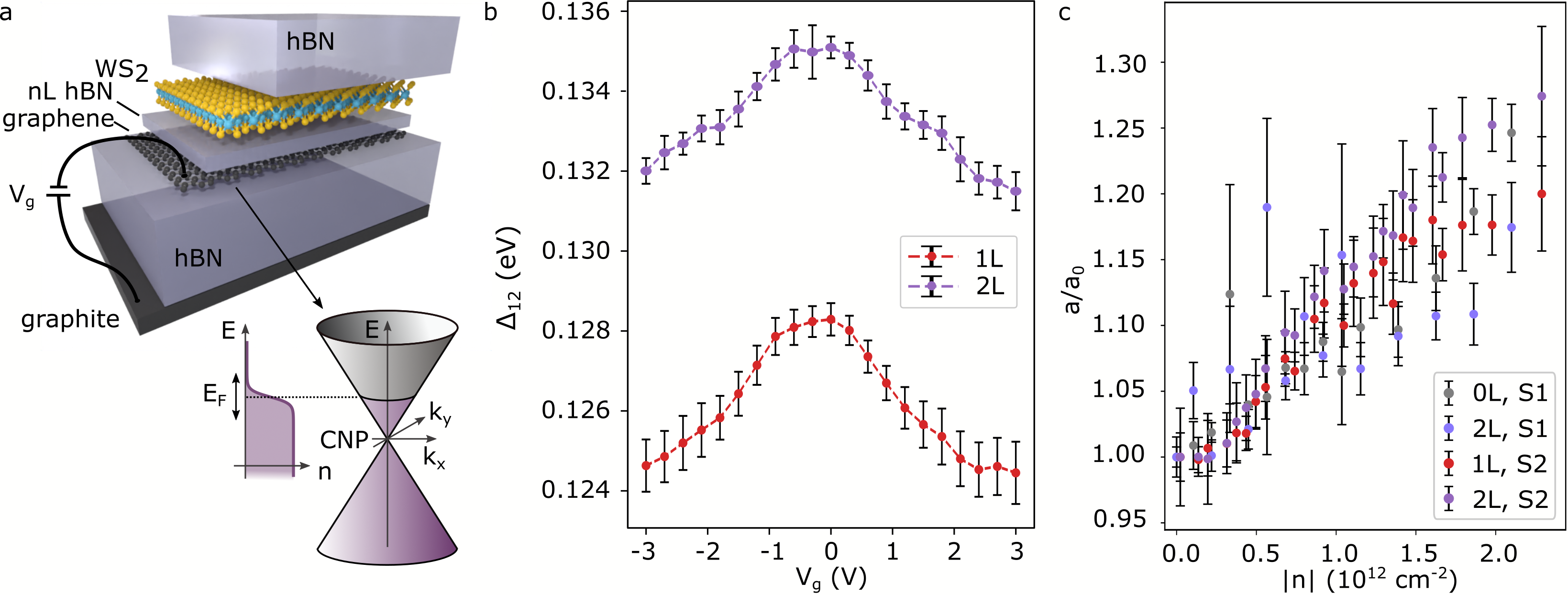}
        \caption{Dependence of the screening on charge carrier density in graphene. a) Sketch of the device geometry (top) used to tune the charge carrier density of graphene from electron ($n>0$) to hole ($n<0$) doping (bottom). b) 1s-2s exciton energy spacing vs backgate voltage $V_\text{G}$ for different hBN spacer layers. c) Relative change of the parameter $a$ (see Eq.~1), proportional to the magnitude of the image charge in graphene, as a function of the absolute charge carrier density $|n|$ of graphene. S1 and S2 denote different samples. Errorbars are standard deviations from analyzing several spatial positions. }
        \label{fig:doping}
    \end{center}
\end{figure*}

We now investigate the possibility to continuously tune the dielectric screening in the same type of heterostructures by modifying the dielectric function of graphene. 
This is achieved by electrostatic doping  \cite{Riis-Jensen2020} via a gate voltage, which tunes the Fermi energy and therefore changes the concentration of free charge carriers in graphene (see Fig. \ref{fig:doping} a). 
At the same time, it modifies its dielectric function  \cite{Hwang2007}.

To quantify the exciton binding energy in the \WS\ layer, reflectance contrast images were taken at various gate voltages. 
From these spectral images, the energies of 1s and 2s excitonic states and their errors are determined in the same way as before, i.e. by analyzing and averaging every viable position in the reflectance contrast image.
To verify that the applied backgate voltages do not significantly change the doping level of the TMD itself, additional photoluminescence measurements were performed, which reveal no signs of trion emission of \WS\ at any gate voltage (see Supplementary Figure 2), possibly due to charge transfer from \WS\ to graphene \cite{Zhu2015, Froehlicher2018}.
We conclude that the doping of the TMD is negligible and the observed modifications of the exciton binding energies result from an external change in dielectric screening.

The energy difference between 1s and 2s states, $\Delta_{12}$, of a sample containing areas of one and two hBN spacer layers is shown as a function of backgate voltage in \fig{fig:doping}b.
For both, positive and negative gate voltage polarities, the measured energy difference decreases with applied voltage. 
This corresponds to the Fermi energy being above and below the charge neutrality point (CNP) of the graphene layer and a population of free electrons or holes, respectively. 
The curve is found to be symmetric, which we ascribe to the electron-hole symmetry of graphene. 
The center is found to be at a voltage of $\approx -0.25$~V, which points to a low residual doping level of graphene. 
The absolute change of $\Delta_{12}$ in \fig{fig:doping}b is larger in areas of smaller interlayer distance, which is also expected from the distance dependence described by equation \ref{eq:dist}. 

We assume that the data can be analyzed in terms of the parameter $a$ as a function of the charge carrier density independent of the carrier type, $a=a(|n|)$.
The data at each interlayer distance and each voltage is therefore scaled by $2(d+d_0)$, with $d_0 = (4.7 \pm 0.3)$ \AA\ being the previously determined interlayer distance between \WS\ and graphene and the factor $2$ stems from the image charge being situated at twice the interlayer distance.

The relative change $a/a_0$, with $a_0=a(|n|=0)$, is shown in \fig{fig:doping}c vs $|n|$ for all samples investigated.
The charge carrier density is calculated from the backgate voltages by $n = \epsilon_0\epsilon_r V_G / d_{\mathrm{hBN}}$ using the bottom hBN thickness $d_{\mathrm{hBN}}$ of the sample, determined using atomic force microscopy and the experimentally determined out-of-plane dielectric constant of hBN $\epsilon_r=3.4\pm0.2$ \cite{Pierret2022}.
Indeed, all investigated sample areas show a similar behavior and no remaining dependence on interlayer distance is discernible, confirming that the charge carrier density of graphene only impacts the material parameter $a$. 
We find $a$ to exhibit a relatively weak dependence on $|n|$ around charge neutrality, followed by a steeper rise which starts to level off at higher charge carrier densities.   
The relatively weak dependence close to the CNP is ascribed to a thermal smearing of the Fermi energy, which leads to a finite density of electron and hole charge carriers $(|n_e| + |n_h|)$ at $|n|=0$ and stands in contrast to experiments at cryogenic temperature \cite{Xu2021}.
Away from the CNP, the dependence is expected to be non-trivial as both the dielectric function of graphene as well as the screening itself depend on both wavevector and frequency and cannot fully be described analytically \cite{Hwang2007, Steinhoff2018}. 
The maximum observed modification of the parameter $a$ is approximately 20\%, which is reached at a charge carrier density slightly above $2\cdot10^{12}$~cm$^{-2}$. 
This shows that, even at room temperature, electrostatic doping of graphene allows to tune Coulomb interactions in nearby 2D semiconductors such as \WS. 

\section{Discussion}

In conclusion, our experimental work quantifies the tunability of excitonic binding energies and the quasiparticle band gap due to dielectric screening in TMD-graphene heterostructures with layer-controlled hBN spacers. 
We find that both are described well by a one-over-distance law. 
The distance at which half of the maximum screening-induced change (for \WS\ in direct contact with graphene) is reached is approximately at two atomic layers of hBN.
This shows that it is possible to simultaneously electrically isolate the two materials \cite{Britnell2012}, while retaining a significant effect of the screening.
It is important to note that, while these results have been obtained at room temperature, the same law is expected to hold at cryogenic temperatures. 

We have furthermore shown that electrostatic gating of the graphene layer leads to a change of the exciton binding energies in \WS\ at various interlayer distances and have quantified the change in a single parameter.
We have reached modifications of this parameter of up to 20\%, limited by a finite density of free charge carriers at charge neutrality and the maximum voltages applied to the samples.

While the tuning of the exciton properties of \WS\ are substantial, we identify several avenues for increasing the absolute changes. 
By using spacer layers and an encapsulating material with an effective dielectric function smaller than that of hBN at frequencies in the infrared \cite{Steinhoff2018}, both the maximum exciton binding energy as well as the absolute changes will be increased.
Furthermore, the gate-induced screening effect at low densities is limited by the minimal charge-carrier density of graphene due to thermal broadening.
While this is straight-forwardly reduced by cooling down the samples, it could also be achieved at room-temperature by replacing graphene with a material possessing a small band gap.

We expect the measured distance dependence of dielectric screening to be valid for various kinds of two-dimensional bilayer structures.
Such systems have recently become of interest in several contexts, such as  for the investigation of interlayer excitons \cite{Fang2014}, doubly charged excitons \cite{Sun2021}, for excitonic Bose-Einstein condensation \cite{Wang2019}, correlation effects \cite{Ma2021} as well as for the investigation of light-matter interactions in TMD superstructures \cite{Kumar2021}.
We have furthermore shown that electrostatically varying the charge carrier density of graphene in the vicinity of a TMD  allows for a continuous and significant change in Coulomb interactions even at room temperature. 
Combining both approaches might lead to a novel way of studying Coulomb interactions in 2D materials showing complex interactions.


\section{Methods}

\subsection{Sample preparation}
Bulk crystals were obtained from hq graphene (\WS), NGS Naturgraphit (graphite) and from NIMS (hBN). 
In the first step, all bulk materials were cleaved using standard tape exfoliation. 
The tapes were pressed onto Si/SiO$_2$ substrates and suitable layers were identified optically by their color contrast to the substrate \cite{Muller2015} (90nm oxide for graphitic layers and thick hBN and 70nm oxide for \WS\ and monolayer hBN).
The correspondence of color contrast and thickness was initially confirmed using atomic force microscopy (for hBN and graphene) and PL spectroscopy for \WS. 

The assembly of the stacks generally follows the recipe described in \cite{Pizzocchero2016}.
The top layer of hBN is picked up with a stamp made of a thin layer of polycarbonate (PC) placed on a piece of polydimethylsiloxane, which is held by a microscope cover glass. 
The other layers are picked up subsequently at approximately from top to bottom. The approximate temperature at the pickup is 90-110 $^\circ$C. 
The whole stack is then released onto a substrate by heating up to 180 $^\circ$C, which releases the polymer film from the PDMS. 
Nine out of ten samples are on Si/SiO$_2$ (285nm) substrates and one stack is on a quartz substrate.
Lastly, the polycarbonate is then dissolved in chloroform, leaving the final stack on the substrate. 

\subsection{Reflectance contrast measurements}
To record the reflectance contrast image, a tungsten whitelight source was focused onto the sample by a 100x, NA$=0.9$ objective.
The sample was then moved using an xy-scanning table and the reflectance spectra were recorded at every position using a grating spectrometer. 
Typical integration times at each pixel are 20s. 

The reflectance contrast $R_C$ was subsequently calculated at each pixel by $R_C = I_{\mathrm{s}}/I_{\mathrm{ref}} -1$, where $I_{\mathrm{s}}$ is the spectrum at the pixel and $I_{\mathrm{ref}}$ is a reference spectrum.
The reference spectrum was obtained by averaging spectra from areas next to the \WS\ sample, in which all other film thicknesses are the same. 
All measurements were performed at room temperature.

\subsection{Fits of the data}
The reflected intensity measured in the experiment is a result of the interference of light reflected at the various interfaces of the samples. 
Each reflection depends on the dielectric function of the materials above and below the interface. 
To fit the measured reflectance contrast spectra, we thus use a thin film interference model (transfer matrix formalism implemented via the python package solcore \cite{Alonso-Alvarez2018}) to calculate the reflectance of the sample region and a reference region (same layer sequence without the sample).
This calculation requires the dielectric function and film thickness of each layer. 
For hBN and graphene as well as silicon and SiO$_2$, reported dielectric functions are used \cite{Lee2019, Weber2010, Green1995, Malitson1965}
The thickness of hBN is determined from optical contrast images and atomic force microscopy measurements.
The thicknesses of graphene and SiO$_2$ are $3.3$\AA\ and 285 nm, respectively, and the thickness of silicon is taken to be semi-infinite as the substrate is several hundred micrometer thick. 

The dielectric function of \WS\ is the quantity which is varied in the fit. 
It is constructed from three Lorentzian oscillators representing the A exciton 1s, 2s and the B exciton 1s absorption features plus a constant offset  $\epsilon(0)$:
$\epsilon_{\mathrm{WS_2}}(\omega) = \epsilon(0) + \sum_{j=1}^3 \frac{f_j}{\omega_j^2 - \omega^2 - i \gamma_j \omega}$.
The energies $\hbar\omega_j$, amplitudes $f_j$ and dampings $\gamma_j$ of these Lorentzians are the three free parameters in the numerical optimization of the calculated vs the measured reflection contrast.
For a better numerical stability of the fit, the derivative of the data is compared to the derivative of the model.

\subsection{Estimation of exciton binding energy}
The exciton energy levels in the two-dimensinal hydrogen model scale like $E_B=\frac1{(n-0.5)^2}$. Therefore, $E_B = \frac98 \cdot \Delta_{12}$.
For a material of non-negligible thickness, the electron-hole potential can be approximated by a Rytova-Keldysh potential \cite{Rytova1967}. 
As a result, the energy-level spacing is non-hydrogenic \cite{Chernikov2014}. 

With increasing external screening, the field-lines of electron-hole interactions are more restricted to the plane of the material. 
This not only decreases the binding energy, but also makes it more 2D hydrogen-like.
Whereas no analytic expressions for $b(\epsilon)$ exists, in Ref. \cite{Waldecker2019}, the band gap as well as 1s and 2s exciton energies are calculated for \WS\ on a substrate of effective dielectric $\epsilon$ (and with vacuum on top). 
Extracting the factor $b=\frac{E_g - E_{\mathrm{1s}}}{E_{\mathrm{2s}}-E_{\mathrm{1s}}}$ from  \cite{Waldecker2019}, we find it to vary between $1.35$ for \WS\ on hBN to $1.2$ for \WS\ on graphene.  
With an extra capping layer of hBN replacing vacuum, $b$ is expected to decrease slightly further, in particular for the less-screaning case (\WS\ on hBN). 
Here, we take a value of $b=1.2$, which is likely in the lower range of possible values. 
Note that this approximation underestimates the change of binding energy between different \WS-graphene distances with an estimated maximum error (0 spacer layers to no graphene) in the 10$\%$ range. 

\section{Data Availability}
The data that support the plots and findings within this paper are available from the corresponding author upon reasonable request.

\section{Acknowledgements}
This project has received funding from the European Union's Horizon 2020 research and innovation programme under grant agreement No 881603 (Graphene Flagship), by the Deutsche Forschungsgemeinschaft (DFG, German Research Foundation) under Germany's Excellence Strategy - Cluster of Excellence Matter and Light for Quantum Computing (ML4Q) EXC 2004/1 - 390534769. K.W. and T.T. acknowledge support from JSPS KAKENHI (Grant Numbers 19H05790, 20H00354 and 21H05233). L.W. acknowledges support by the Alexander von Humboldt Foundation. 

\section{Author contributions}
L.W. conceived the project. 
K.W. and T.T. grew hBN crystals.
D.T. and M.S. produced the samples and conducted the measurements. 
D.T, M.S. and L.W. analyzed the data, which was discussed by all authors.
L.W. and D.T. wrote the manuscript with input from all authors. 

\bibliography{Tebbe_screening_WS2}

\section{Competing Interests}
The authors declare no competing interests.

\end{document}